\documentclass[12pt,preprint]{aastex}

\slugcomment{Accepted for publication in ApJ Letters.}
\shorttitle{UV Detection of the Tidal Disruption of a Star}
\shortauthors{Gezari et al.}

\newcommand {\apgt} {\ {\raise-.5ex\hbox{$\buildrel>\over\sim$}}\ }
\newcommand {\aplt} {\ {\raise-.5ex\hbox{$\buildrel<\over\sim$}}\ }

\newcommand{\mbh}{M_{\rm BH}}%
\newcommand{\msun}{M_{\sun}}%
\newcommand{\etal}{et al.~}%
\begin{document}

\title{Ultraviolet Detection of the Tidal Disruption of a Star\\by a Supermassive Black Hole}

\author{S.~Gezari\altaffilmark{1},
 D.~C.~Martin\altaffilmark{1},
  B.~Milliard\altaffilmark{2},
  S.~Basa\altaffilmark{2}, 
  J.~P.~Halpern\altaffilmark{3},
  K.~Forster\altaffilmark{1},
  P.~G.~Friedman\altaffilmark{1},
  P.~Morrissey\altaffilmark{1},
  S.~G.~Neff\altaffilmark{4},
  D.~Schiminovich\altaffilmark{3},
  M.~Seibert\altaffilmark{1}, 
  T.~Small\altaffilmark{1}, and
  T.~K.~Wyder\altaffilmark{1}
}

\altaffiltext{1}{California Institute of Technology, 
        MC 405-47,
        1200 East California Boulevard,
        Pasadena, CA  91125 \email{suvi@srl.caltech.edu}}

\altaffiltext{2}{Laboratoire d'Astrophysique de Marseille, 
        13376 Marseille Cedex 12, France}

\altaffiltext{3}{Department of Astronomy, 
    Columbia University, 
        New York, NY  10027}

\altaffiltext{4}{Laboratory for Astronomy and Solar Physics, 
     NASA Goddard Space Flight Center, 
      Greenbelt, MD  20771}

\begin{abstract}
A supermassive black hole in the nucleus of a galaxy will be
revealed when a star passes close enough to be torn
apart by tidal forces and  
a flare of radiation is emitted by the stream 
of stellar debris that plunges into the black hole.  
Since common active galactic nuclei have accreting black holes
that can also produce flares, a
convincing demonstration that a stellar tidal disruption has occurred generally begins
with a ``normal'' galaxy that has no evidence of prior nuclear activity.
Here we report a luminous UV flare from an elliptical galaxy at $z = 0.37$ in the Groth
field of the \textsl{GALEX} Deep Imaging Survey that has no evidence of a Seyfert nucleus from optical spectroscopy and X-ray imaging obtained during the flare.  Multiwavelength data collected at the time of the event, and for 2 years following, allow us to constrain, for the first time, the 
spectral energy distribution 
of a candidate tidal disruption flare from optical through X-rays.
The luminosity and temperature of the radiation and the decay curve of the flare are in excellent
agreement with theoretical predictions for the tidal disruption of a star, and provide the strongest empirical evidence for a stellar disruption event to date.  
\end{abstract}

\keywords{galaxies: nuclei --- Ultraviolet: galaxies}

\section{Introduction \label{intro}}

Dynamical studies of nearby galaxies show that all galaxies with a bulge host a
supermassive black hole in their nucleus \citep{mag98}.  
Stellar dynamical models \citep{mt99,wm04} 
predict that once every 10$^4$--10$^{5}$ yr the orbit of a star in the nucleus
of a galaxy will pass within
the tidal disruption radius of its central supermassive black hole, $R_{T} \approx R_{\star}(\mbh/M_{\star})^{1/3}$,
and the star will be disrupted.  
A stellar disruption results in a flare
of electromagnetic radiation \citep{fr76,lid79} as the bound fraction of the stellar debris,
\aplt $0.5 M_{\star}$ (Rees 1988; Ayal \etal 2000), falls back onto the black hole
and is accreted.  There is a critical
black hole mass above which $R_{T}$ is smaller than the Schwarzschild radius ($R_{s}$),
and the star is swallowed whole without disruption \citep{hills75}.  For a solar-type star, $M_{crit} \sim 10^{8} \msun$.
The luminosity and decay of a tidal disruption flare is dependent on the mass and spin
of the central black hole, and can be used to directly probe dormant black holes in
galaxies for which the sphere of influence of the black hole ($R_{sph}=G\mbh/\sigma_{\star}^2$) is unresolved, 
and a dynamical measurement of the black hole mass is not possible.

The start of the flare, $t_{0}$, occurs when the most tightly bound debris first returns to pericenter ($R_{p}$) 
after the star is
disrupted at time t$_{D}$, where 

($t_{0}-t_{D}) = 0.11 k^{-3/2} (R_{p}/R_{T})^3 
(R_{\star}/R_{\odot})^{3/2} (M_{\star}/M_{\odot})^{-1} (\mbh/10^{6} \msun)^{1/2}$ yr 

\noindent (Li \etal 2002),
and $k = 1$ for a star with no spin, and $k = 3$ for a star that is spun up by tidal interactions
to near break-up on disruption.  After $t_{0}$, the return rate of the bound debris to pericenter declines as function of time as
$\dot{M}(t) \propto [(t-t_{D})/(t_{0}-t_{D})]^{-5/3}$ \citep{evans89}, which determines the
characteristic power-law decay of the luminosity of the flare over the following months and years.
When the mass accretion rate is super-Eddington, 

$\dot{M}_{Edd} \equiv L_{Edd}/(\epsilon c^{2}) = 
0.025 (\mbh/10^{6} \msun) (\epsilon/0.1)^{-1} \msun {\rm yr}^{-1}$, 

\noindent where $\epsilon$ is the efficiency of converting mass into radiation, 
the debris is accreted in a thick disk that radiates close to the Eddington luminosity \citep{ulmer99}, 
$L_{Edd} =  1.3 \times 10^{44} (\mbh/10^{6} \msun)$ ergs s$^{-1}$.  
The spectrum of a tidal disruption event can thus be characterized 
by the blackbody temperature
of a thick disk radiating at $L_{Edd}$ at radii ranging from $R_{T}$ to $R_{p}$, $T_{eff} \approx
(L_{Edd}/4\pi\sigma R^{2})^{1/4}$, which for a solar-type star is 
$\sim (2-5) \times 10^{5}$ K, and peaks in the extreme-UV \citep{ulmer99}.

The most unambiguous cases for a stellar disruption occur from host galaxies with no evidence of an active galactic nucleus (AGN) for which an upward fluctuation in the accretion rate could also explain a luminous UV/X-ray flare.  A UV flare from the nucleus of the elliptical galaxy NGC 4552 was proposed to be the result of the tidal stripping of a stellar atmosphere \citep{ren95}; however the possible presence of a persistent, low-luminosity AGN detected in hard X-rays (Xu \etal 2005) makes this interpretation uncertain.  The \textsl{ROSAT} All-Sky survey conducted in 1990--1991
sampled hundreds of thousands of galaxies in the soft X-ray band, and detected luminous ($10^{42}-10^{44}$ ergs s$^{-1}$),
soft [$T_{\rm bb} = (6 - 12) \times 10^{5}$ K] X-ray flares 
from several galaxies with no previous evidence
for AGN activity, and with a flare rate of $\approx 1 \times 10^{-5}$ yr$^{-1}$ per galaxy \citep{donley02}, that is consistent with
the theoretical stellar disruption rate.
A decade later, follow-up \textsl{Chandra} and \textsl{XMM-Newton} observations of three of the galaxies demonstrated that they had
faded by factors of $240 - 6000$, consistent 
with the ($t-t_{D})^{-5/3}$ decay of a tidal disruption flare (Komossa \etal 2004; Halpern \etal 2004).  Follow-up \textsl{Hubble Space Telescope} (\textsl{HST}) Space Telescope Imaging
Spectrograph narrow-slit
spectroscopy confirmed two of the galaxies as inactive, qualifying them as the 
most convincing hosts of a tidal disruption event \citep{gez03}.  The \textsl{ROSAT} flare with
the best sampled light curve was successfully modeled as the tidal disruption of a brown dwarf or planet (Li \etal 2002), although its host galaxy was subsequently found to have a low-luminosity Seyfert nucleus \citep{gez03}.

\section{Observations \label{obs}}

We initiated a program to take advantage of the UV sensitivity,
large volume, and temporal sampling of the \textsl{Galaxy Evolution Explorer} (\textsl{GALEX}) Deep Imaging Survey (DIS) 
to search for stellar disruptions in the nuclei 
of galaxies over a large range of redshifts.
The DIS covers 80 deg$^{2}$ of sky in the far-ultraviolet (FUV; $\lambda = 1344-1786 $\AA) and near-ultraviolet (NUV; $\lambda = 1771-2831 $\AA) with a total exposure time of 30--150 ks, 
that is accumulated in $\sim 1.5$ ks eclipses (when the satellite's 98.6 minute 
orbit is in the shadow of the Earth). Due to target visibility
and mission planning constraints, some DIS fields are observed over a baseline of 2--4
years to complete the total exposure time.  This large range in cadence of the observations
allows us to probe variability on timescales from hours to years.
Here we present a UV flare discovered in the \textsl{GALEX} 1.2 deg$^{2}$ Groth field
at a position of right ascension $14^{\rm h}19^{\rm m}29.^{\rm s}8$, declination $+52^\circ 52\arcmin 6\arcsec$.  We use multiwavelength survey data from the All-Wavelength Extended Groth Strip
International Survey (AEGIS) \citep{dav06} and the Canada-France-Hawaii Telescope 
Legacy Survey (CFHTLS) to identify the host of the flare, and 
study the broadband properties of the flare and its subsequent decay.
Throughout this Letter, calculations are made using 
\textsl {Wilkinson Microwave Anisotropy Probe} cosmological parameters \citep{ben03}: 
$H_{0}$ = 71 km s$^{-1}$ Mpc$^{-1}$, $\Omega_{M} = 0.27$, and $\Omega_{\Lambda} = 0.73$.

\begin{figure}
\epsscale{.7}
\plotone{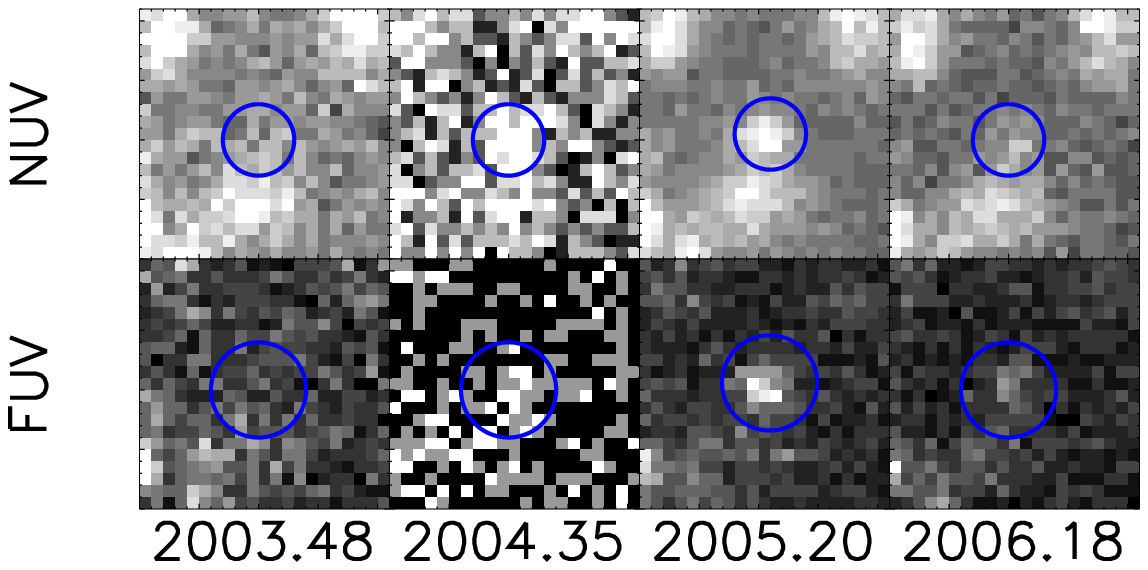}
\caption{\textsl{GALEX} images of the flaring source over 4 years of DIS eclipses coadded
into yearly epochs of NUV ($t_{exp}$= 16, 1.5, 110, and 29 ks) and FUV ($t_{exp}$= 16, 1.5, 41, and 29 ks)
observations, labeled with the mean date of each coadd.  Blue circles show the 9$\arcsec$ diameter aperture in the NUV, and the 12$\arcsec$ diameter aperture in the FUV used for photometry.
No source is detected in NUV and FUV in the 2003.48 epoch.  
A smaller aperture was used in the NUV to avoid contamination by a pair of foreground interacting galaxies detected
in the NUV just South of the source.  The counts from the background in the aperture 
were determined from the pipeline background image, and
an aperture-correction factor of 80\% and 85\% was applied to the NUV and FUV aperture fluxes respectively, to 
recover the total energy enclosed in a point source. \label{fig:fl}}
\end{figure}

\begin{figure}
\epsscale{.7}
\plotone{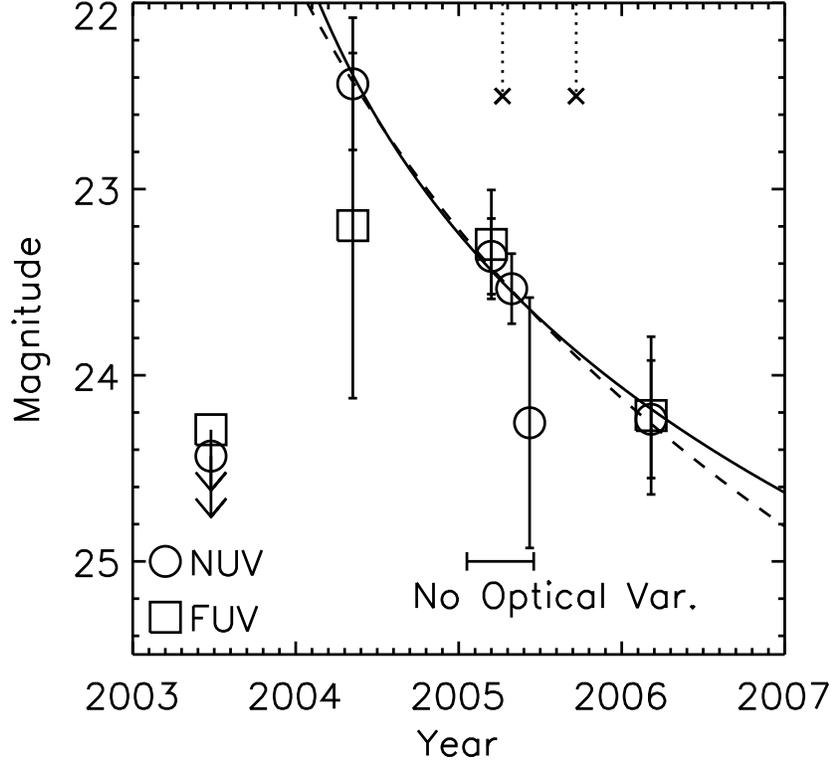}
\caption{\textsl{GALEX} light curve in the FUV ($\lambda_{eff} = 1528 $\AA) and NUV ($\lambda_{eff} = 2271 $\AA).  
Error bars show 68\% confidence intervals and arrows show 95\% confidence upper limits
based on  Bayesian statistics.  For the later
observations in 2005 the FUV detector was temporarily not operational. 
The least-squares fit to the NUV light curve with a ($t-t_{D})^{-5/3}$ decay
is shown with a thick solid line, yielding a best fit time of disruption $t_D$ = 2003.3, with a 1 $\sigma$ error of 0.2 yr.   The
least-squares fit to the NUV light curve with the power-law index, ($t-t_{D})^{-n}$, and $t_{D}$ allowed to vary is shown with a 
dashed line, with $t_{D} = 2002 \pm 2$ and $n = 3 \pm  2$.  Dotted lines
with an X indicate times of two sets of 100 ks \textsl{Chandra} 0.3 -- 10 keV observations.  An extremely soft X-ray source was detected during the first set observations, and no source was detected in the second set of observations.  CFHTLS monitoring data from 2005 January to June
in $g~(\lambda_{eff} = 4763$\AA)$, r~(\lambda_{eff} = 6174$\AA)$, 
i~(\lambda_{eff} = 7619$\AA), and $z~(\lambda_{eff} = 8847$\AA) with $m_{lim}$ = 25, detect
no variable optical source during the UV flare. \label{fig:lc}}
\end{figure}

The UV flare reported here was discovered as a result of its large amplitude of variability
during 4 years of DIS observations.  Figure \ref{fig:fl} shows
the discovery images in the NUV and FUV of the source, with the eclipses co-added into yearly
epochs, and
Figure \ref{fig:lc} shows its detailed NUV+FUV light curve.    The source
is undetected in the NUV and FUV in 2003 June 21 -- 29, and then appears in 2004 March 25 -- June 24, 
indicating an amplitude of variability
of $\Delta m$ \apgt 2 mag.  The source then decays monotonically by $\sim$ 2.0 mag over 2 years 
to almost below the detection threshold
in 2006 March 5 -- 7. 
Archival AEGIS 50 ks \textsl{Chandra} (0.3 -- 10 keV) observations on 2005 April 6 and 7 detected 
an extremely soft X-ray source that appeared on the second day of the observations, with 10 photons 
with energies between 0.3 and 0.8 keV, indicating a spectral slope, $f_{E} \propto E^{-\Gamma}$,
of $\Gamma = 7 \pm 2$ (reduced $\chi^{2}=0.92)$, 
for a column density of neutral hydrogen fixed to the Galactic value. 
No source was detected in the following 50 ks \textsl{Chandra} observations on 2005 September 20 and 23.   

\section{Interpretation \label{int}}

\begin{figure}
\epsscale{.7}
\plotone{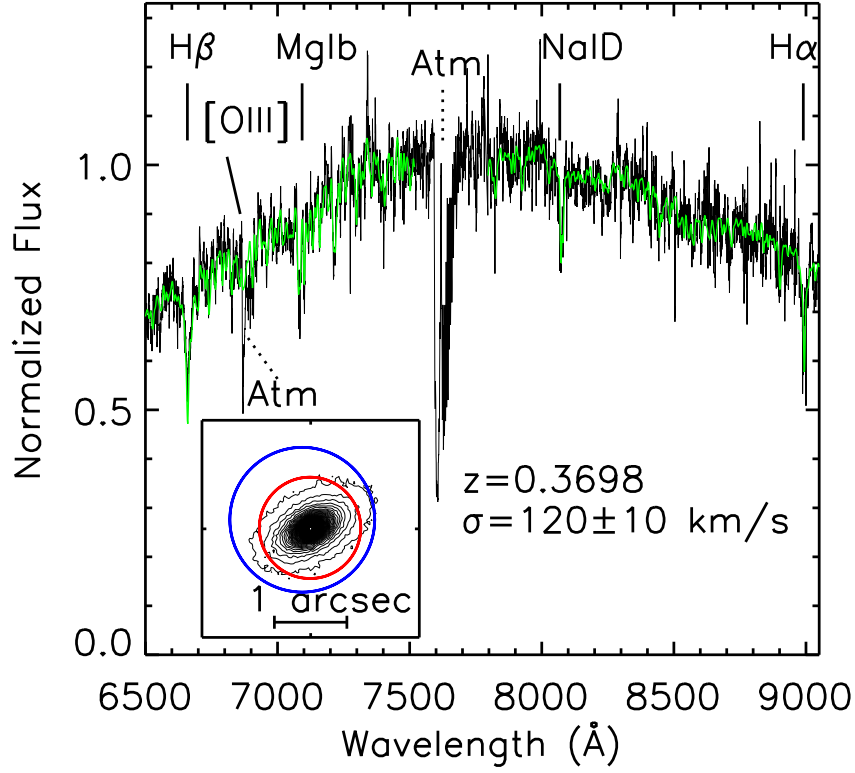}
\caption{{\it Image:} AEGIS \textsl{HST} ACS {\it I}-band image ($\lambda_{eff} = 8140 $\AA) of the galaxy host with the \textsl{GALEX} 
1$\arcsec$ position error circle ({\it blue}) of the UV flare and the \textsl{Chandra} 0.$\arcsec$7 
position error circle ({\it red}) of the soft X-ray source,
with a 1$\arcsec$ systematic astrometry correction applied to the \textsl{Chandra} position measured from the
shift of the \textsl{Chandra} astrometry with respect to ACS for a nearby star.  The morphology of the galaxy is determined from a two-dimensional
 bulge/disk
decomposition of the ACS I-band image \citep{sim98}, which results in a bulge fraction of 0.72, and a half-light radius
of the bulge, $R_{e}$ = 0.$\arcsec$43.  {\it Spectrum:} AEGIS Keck DEIMOS spectrum from the DEEP2 survey of the galaxy host obtained on 2005 March 9, 
smoothed by 6 pixels ($\sim$ 2 \AA), and not corrected for the instrumental response.  The spectrum shows strong stellar absorption
lines typical of an early-type galaxy, with a marginal detection of narrow [O~III] $\lambda 5007$ line emission.  The best-fitting Bruzual-Charlot early-type galaxy template \citep{bruz03} scaled to the continuum of the
spectrum  is shown in green.  The luminosity-weighted line of sight velocity dispersion 
measured using stellar templates \cite{mo05} within the 1$\arcsec$ slit width of DEIMOS is 120 $\pm$ 10 km/s, which corresponds to an aperture of 1.2$R_{e}$, for which the
correction to an aperture of $R_{e}$ is $<$ 5\% \citep{geb00}. \label{fig:acs}}
\end{figure}

The UV flare and variable soft X-ray source are coincident with an elliptical galaxy at $z$ = 0.3698 (luminosity distance, $d_{L} = 1970$ Mpc).  
Figure \ref{fig:acs} shows the AEGIS {\textsl HST} ACS optical image with the UV and X-ray positions overplotted, and  
the AEGIS Keck DEIMOS optical spectrum from the DEEP2 survey of the host galaxy
taken on 2005 March 9.
The spectrum has strong stellar absorption features typical of an early-type galaxy.
We marginally 
detect faint narrow [O~III] $\lambda$5007 line
emission with $L$([O~III]) = 
$8 \pm 4 \times 10^{39}$ ergs s$^{-1}$.
The non-detection of a hard X-ray source by \textsl{Chandra} places an upper limit to the 3--20 keV 
luminosity of a power-law AGN spectrum in the nucleus of $L_{X}$ \aplt 1 $\times$ $10^{42}$ ergs s$^{-1}$.  
The correlation between $L$([O~III]) and $L_{X}$(3--20 keV) observed for local AGNs \citep{heck05} predicts 
$L$([O~III]) $< 10^{39}$ ergs s$^{-1}$, which suggests that the possible [O~III] line emission has too high an $L$([O~III])/L$_{X}$ ratio to be associated with AGN activity, and may be from low levels
of star formation.
We obtained a follow-up spectrum
after the flare discovery on 2006 June 1 with the Double Spectrograph on the Palomar 200 inch telescope, and again detected no 
strong emission lines.
CFHTLS variability monitoring data from 2005 January to June 
show no detection of a variable optical source in the galaxy, with up to 5 observations per month with a sensitivity of $m_{lim} = 25$.  

The lack of variable optical
emission during the UV flare puts an important constraint on its nature.  Supernovae (SNe) have light curves that extend over months, 
similar to the flare observed here, but their emission peaks at optical wavelengths, and is intrinsically faint shortward of $\sim$ 3000 \AA~\citep{pan03,brown05}.
Gamma ray bursts (GRBs) have power-law decays, but again, their emission is brighter at optical 
wavelengths \citep{mes97}. 
The characteristic power-law continuum of AGN emission extends
from optical to extreme-UV as $f_{\nu} \propto \nu^{-\alpha}$, where 0.43 $\le \alpha \le$ 2
\citep{zheng97,diet02}.  
If the UV flare were an upward fluctuation of AGN continuum emission, 
a power-law spectrum with $\alpha \ge 0.43$ would produce an optical source with $m <$ 23 mag during 
the CFHTLS observations and power a broad H$\beta$ line with a luminosity of 
\apgt 1 $\times~10^{41}$ ergs s$^{-1}$ \citep{diet02} at the time of the Keck spectrum,
both of which are definitively ruled out by the observations.  In summary, the flare is most
probably not due to AGN activity because it displays (1) no optical variability, (2) no broad emission lines, (3) a soft
X-ray spectrum with no hard X-ray component, and (4) an early-type galaxy spectrum with possible
narrow [O~III] emission with a high [O~III]/L$_{X}$ ratio characteristic of star formation.
In addition, a SN or GRB origin is unlikely because of the high UV/optical ratio.

\begin{figure}		
\epsscale{.7}
\plotone{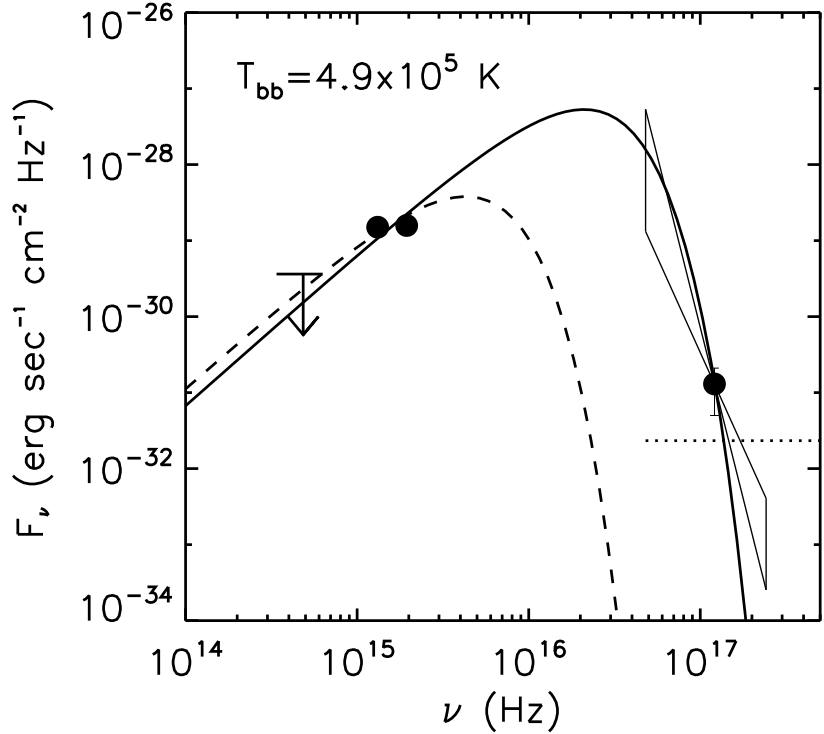}
\caption{Spectral energy distribution from optical to X-rays of the flare on 2005 April 7, with the best
blackbody fit with $T_{bb}=4.9\times$10$^5$ K shown with a solid line.   The spectral slope of the extremely soft X-ray detection, $f_{\nu} \propto E^{-\Gamma}$, 
of $\Gamma = 7 \pm 2$ is also shown, corresponding to an flux density at 0.5 keV of (1.3 $\pm 0.8) \times 10^{-31}$ ergs s$^{-1}$ cm$^{-2}$ Hz$^{-1}$,
a factor of $\sim$ 4 above the detection limit of \textsl{Chandra} ({\it dotted line}).
Since the X-ray flux appears above the detection threshold of \textsl{Chandra} 
for only part of the observations, the blackbody fit gives an upper limit
to $T_{bb}$ during the flare.
Dashed line shows the lowest temperature blackbody consistent with the UV flux densities and optical upper limits, with $T_{bb}=1\times$10$^5$ K. \label{fig:sed}}
\end{figure}

The spectral energy distribution (SED) 
of the flare from optical to X-rays on 2005 April 7 is shown in Figure \ref{fig:sed}.  A blackbody fit to the SED is strongly
constrained by the soft X-ray flux density, which is best fit for a rest-frame $T_{bb}= 4.9 \times 10^{5}$ K, shown with a 
solid line in Figure \ref{fig:sed}, which for $L_{\nu_{e}=(1+z)\nu_{o}}=  f_{\nu_{o}} (4\pi d_{L}^2)/(1+z)$, 
corresponds to $R_{bb} = 6.2 \times 10^{12}$ cm and $L_{bol} = 1.2 \times 10^{45}$ ergs s$^{-1}$.  The steep Wein's tail of the
blackbody curve is consistent with the extremely soft spectral slope of the X-ray source.  
However, since the X-ray flux varies on a short timescale, and only appears above the detection threshold for part of the observations, 
we consider this fit an upper limit to $T_{bb}$ for the duration of the flare. 
The lowest $T_{bb}$ that is consistent with the UV flux densities and the optical upper limits is $\sim$ 1 $\times~10^{5}$ K, shown with a 
dashed line in Figure \ref{fig:sed}, which corresponds to  $R_{bb}=1.6 \times 10^{13}$ cm and $L_{bol} = 1.3 \times 10^{43}$ ergs s$^{-1}$.
This range of temperatures and luminosities is in excellent agreement with the theoretical predictions for a tidal disruption flare.

\section{Discussion \label{disc}}

The characteristic tidal disruption ($t-t_{D}$)$^{-5/3}$ power-law decay 
fit to the UV light curve yields a best-fit time of disruption of $t_{D}$=2003.3 $\pm$ 0.2.  If we allow the
power-law index of the decay, ($t-t_{D}$)$^{-n}$, to vary as well as $t_{D}$, 
we get a comparable fit, but with much less constrained values for $t_{D} = 2002 \pm 2$ and $n = 3 \pm 2$.
The \textsl{GALEX} observations do not cover
the rise of the UV flux to its peak value, however we can constrain the time of the peak (t$_{0}$) to be between the
time of the \textsl{GALEX} detection beginning on 2004 March 25 and the latest date of the non-detection on 2003 June 29,
($t_{0}-t_{D}) = 0.2 (\pm 0.2) - 0.9 (\pm 0.2)$ yr.
We estimate the total energy released during
the flare by integrating the best-fit ($t-t_{D}$)$^{-5/3}$ light curve 
from the time of the peak of the flare to infinity,
$E_{tot} = \int_{t_{0}}^{\infty}{L(t)dt}$.  If on 2005 April 6--7, $1.3 \times 10^{43}$ ergs s$^{-1} \aplt L_{bol} \aplt 1.2 \times 10^{45}$ ergs s$^{-1}$,
and ($t_{0} - t_{D}$) $\aplt$ $1.1$ yr, this yields a minimum total energy 
released of $1.8 \times 10^{51}$ ergs $\aplt E_{tot} \aplt 1.7 \times 10^{53}$ ergs,
and a minimum mass accreted, $M_{acc} = E_{tot}/(\epsilon c^2)$, of  $0.01 \msun (\epsilon/0.1)^{-1} \aplt M_{acc} \aplt 0.9 \msun (\epsilon/0.1)^{-1}$, assuming $T_{bb} = (1-5) \times 10^{5}$ K during the flare.   

For a solar-type
star, the limits on ($t_{0} - t_{D}$) imply $k^{3} 3 \times 10^{6} \msun \aplt \mbh \aplt  k^{3} 7 \times 10^{7} \msun$, where $k$ depends on the spin-up of the star, with a hard
upper limit on the black hole mass of $\mbh < M_{crit} = 1.15 \times 10^{8} (r^3/m )^{1/2} \msun$, where $r=R_{\star}/R_{\odot}$, and $m=M_{\star}/M_{\odot}$, in order to satisfy the condition that $R_{T} > R_{s}$, and a stellar disruption can occur.
The radius of the flare emission also places constraints on $\mbh$.  The debris disk should form within the tidal disruption radius of the central black hole.  The blackbody fits to the flare SED have $R_{bb} \aplt 1.6 \times 10^{13}$ cm which, if $R_{bb} \aplt R_{T}$, places a lower limit on the black hole mass of $\mbh \apgt 1.2 \times 10^{7} \msun$.  The radius of the emission should also be greater than the minimum stable particle orbit for the black hole ($R_{ms}$), which ranges from $R_{ms} = 6R_{g}$ for a black hole with no spin, down to $R_{ms}$ = $R_{g}$ for a maximally spinning black hole, where R$_{g} = G\mbh/c^{2}$.  For $R_{bb} > R_{ms}$, this requires that $\mbh < 1.1 \times 10^{8} \msun$, and is consistent with $\mbh < M_{crit}$ for a solar-type star.
We can also use the strong correlation between black hole mass and stellar 
velocity dispersion established for local galaxies \citep{tr02}, log($\mbh/\msun) \sim 8.13 + 4.02{\rm log}(\sigma_e/200$ km s$^{-1}$), to
estimate $\mbh$.
The stellar velocity dispersion within the half-light radius ($R_{e}$) of the bulge of the galaxy, $\sigma_{e} = 120 \pm 10$ km s$^{-1}$, yields a mass of $2^{+2}_{-1} \times 10^7 \msun$,
where the 1$\sigma$ error includes the intrisic scatter of the correlation of $\sim$ 0.3 dex.  

To summarize, the black hole mass estimated from the $\mbh-\sigma$ relation ($\mbh = 2^{+2}_{-1} \times 10^7 \msun$) is consistent with the constraints on the black hole mass from the blackbody spectrum such that $R_{ms} < R_{bb} \aplt R_{T}$ ($1.2 \times 10^{7} \msun \aplt \mbh < 1.1 \times 10^{8} \msun$), with the limits derived from the time delay between the time of the disruption and the peak of the flare ($k^{3} 3 \times 10^{6} \msun \aplt \mbh \aplt  k^{3} 7 \times 10^{7} \msun$), and with
the condition that  $\mbh < M_{crit} = 1.15 \times 10^{8}$ for a solar-type star to be disrupted.  It is encouraging that these various {\it independent} black hole mass estimates are in agreement with each other as long as $k < 3$, and the star was not spun-up to near break-up on disruption. 
The \textsl{GALEX} DIS has proven to be a promising survey for detecting stellar disruption events so far.  
With a larger sample of tidal disruption flares, we can begin to measure the distribution
of masses and spins of black holes in the nuclei of normal galaxies as a function of redshift, 
unbiased by the minority of galaxies that are AGNs.

\acknowledgements
We thank the anonymous referee for helpful comments, including the suggestion to add a discussion of the radius of the flare emission.  We thank C.~L. Slesnick for carrying out our target-of-opportunity observation with the Double Spectrograph
on the Palomar 200 inch telescope,
V. Villar for the two-dimensional bulge/disk composition of the AEGIS \textsl{HST} ACS image, and S.~M. Moran
for measurement of the stellar velocity dispersion of the AEGIS DEEP2 DEIMOS spectrum.
S.G. was supported in part by the Volontariat International-CNES of France, and through \textsl{Chandra} Grant Award G06-7099X
issued by the \textsl{Chandra X-ray Observatory}, 
which is operated by the SAO 
for and on behalf of NASA.
We gratefully acknowledge NASA's support for construction, 
operation, and science analysis for the \textsl{GALEX} mission, 
developed in cooperation with CNES and the Korean Ministry of Science and Technology.
The AEGIS collaboration acknowledges support from the NASA/ESA \textsl{HST}
grant GO-10134 for the Extended Groth Strip observations, obtained at STScI, which is operated by AURA, Inc. under a NASA contract.
The AEGIS collaboration also acknowledges support from the NSF 
grant AST-0507483 for the DEEP2 survey observations with
DEIMOS at the W. M. Keck Observatory. 
Based on observations obtained with MegaPrime/MegaCam, a joint
project of CFHT and CEA/DAPNIA, at the
CFHT which is operated by the NRC of
Canada, the CNRS of France, and the
University of Hawaii. This work is based in part on data products
produced at TERAPIX and the Canadian Astronomy Data Centre as part of
the CFHT Legacy Survey, a collaborative
project of NRC and CNRS.

\clearpage

\clearpage

\clearpage

\clearpage

\end{document}